# A distributed service for on demand end to end optical circuits

**Ramiro Voicu\*, Iosif Legrand\*, Harvey Newman\*,
Nicolae Tapus\*\*, Ciprian Dobre\*\***

*California Institute of Technology, Pasadena, California, USA*
*(e-mail: Ramiro.Voicu@cern,ch, Iosif.Legrand@cern.ch, Harvey.Newman@cern.ch)*
*\*\*Unversity Politehnica of Bucharest, Splaiul Independentei 313, Bucharest, ROMANIA*
*(e-mail:ntapus@cs.pub.ro, ciprian.dobre@cs.pub.ro)*

**Abstract:** In this paper we present a system for monitoring and controlling dynamic network circuits inside the USLHCNet network. This distributed service system provides in near real-time complete topological information for all the circuits, resource allocation and usage, accounting, detects automatically failures in the links and network equipment, generate alarms and has the functionality to take automatic actions. The system is developed based on the MonALISA framework, which provides a robust monitoring and controlling service oriented architecture, with no single points of failure.

## 1. INTRODUCTION

Over the last few years the ever-increasing demand for data services has been the driving force behind the wide deployment of new technologies to increase network capacity.

The High Energy Physics experiments at CERN's Large Hadron Collider (1)(2) will face unprecedented challenges due to the volume and complexity of the experimental data, and the need for collaboration among scientists located around the world. The massive datasets which will be acquired, processed, distributed and analyzed are expected to grow to the 100 Petabyte level and beyond by 2010. Managing these large amounts of data requires a different approach for data transfers from both application and network perspective. The latest developments in high speed networks investigate the idea of hybrid networks with both type of packet switched (traditional) and circuit switched segments.

This idea is actively investigated by projects such as DRAC (3), DRAGON (4), or OSCARS (5). However, their focus is only on dynamic network configuration for the purpose of bandwidth time-slots reservations.

We used the MonALISA (6)(7) (Monitoring Agents using a Large Integrated Service Architecture) framework for developing a system implemented as a distributed set of collaborating agents. The MonALISA system provides near real-time access to complete monitoring information and allows analyzing and processing this information in a dynamic network of agents, capable to build higher level services.

The system has been developed, tested and is part of it currently used in the USLHCNet (8) network, which provides the access between CERN and major US Tier1 sites: BNL (9) and Fermilab (10). The network infrastructure is hybrid having both packet switched devices (routers) and circuit-oriented devices at both layer one (Glimmerglass (11) and Calient (12) optical switches) and layer two (Ciena Core Director (13)). The system we present in this paper was designed to provide a distributed network service for other higher level applications or clients.

## 2. THE MonALISA FRAMEWORK

MonALISA (Monitoring Agents in A Large Integrated Services Architecture) provides a distributed service for monitoring, control and global optimization of complex systems. It is based on an ensemble of autonomous multi-threaded, agent-based subsystems which able to collaborate and cooperate to perform a wide range of monitoring and decision tasks in large scale distributed applications, and to be discovered and used by other services or clients that require such information. It is a fully distributed system with no single point of failure. The scalability of the system derives from the use of a multi threaded engine to host a variety of loosely coupled self-describing dynamic services, the ability of each service to register itself and then to be discovered and used by any other services, or clients that require such information.

*2.1 Communication between agents inside the MonALISA framework*

The MonALISA architecture, presented in Figure 1, is based on four layers of global services. The network of JINI (14) - Lookup Discovery Services (LUS) provides dynamic registration and discovery for all other services and agents. MonALISA services are able to discover each other in the distributed environment and to be discovered by the interested clients. The registration uses a lease mechanism. If a service fails to renew its lease, it is removed from the LUSs and a notification is sent to all the services or other application that subscribed for such events. Remote event notification is used in this way to get a real overview of this dynamic system.

The second layer, represents the network of MonALISA service that host many monitoring tasks through the use of a multithreaded execution engine and to host a variety of loosely coupled agents that analyze the collected information in real time. The collected information can be stored locally in databases. Agents are able to process information locally and to communicate with other services or agents to perform global optimization tasks. A service in the MonALISA framework is a component that interacts autonomously with other services either through dynamic proxies or via agents that use self-describing protocols. By using the network of lookup services, a distributed services registry, and the discovery and notification mechanisms, the services are able to access each other seamlessly. The use of dynamic remote event subscription allows a service to register an interest in a selected set of event types, even in the absence of a notification provider at registration time. The third layer of Proxy services, shown in the figure, provides an intelligent multiplexing of the information requested by the clients or other services and is used for reliable communication between agents. It can also be used as an Access Control Enforcement layer to provide secures access to the collected information.

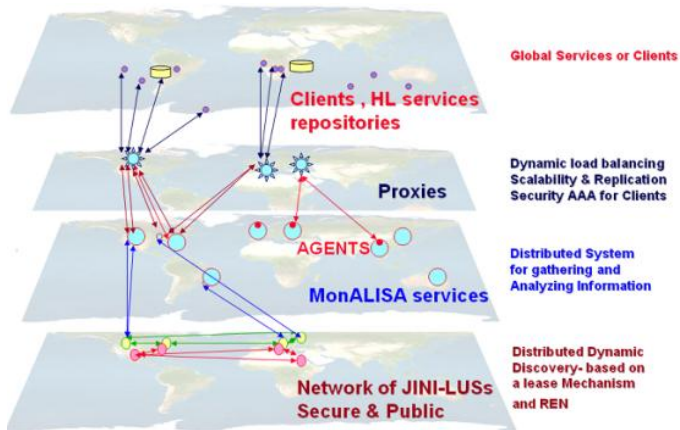

Fig. 1 The MonALISA Architecture

Higher level services and client access the collected information using the proxy layer of services. A load balancing mechanism is used to allocate these services dynamically to the best proxy service.

The ability of the agent-based architecture to self-organize, adapt rapidly to changing conditions, and be invested with increasing degrees of intelligence and autonomy, dramatically simplifies the operation and management of global-scale grids.

Inside the MonALISA framework the services are able to communicate between them through a set of proxy services. The system uses Jini for dynamic registration and discovery mechanism of all the services and agents. The communication between services is made through a set of proxy services which provides an intelligent way multiplexing the information requested by other services or clients. The proxy and lookup services are replicated to guarantee reliability through redundancy. Every proxy can handle more than 1000 messages per second and the framework uses more than one proxy to achieve reliable communication between agents. Agents are hosted by the monitoring services and are able to publish and discover themselves based on a group name. Every agent must be a member of at least one group and is able to communicate only with the agents from the same group. The messages sent between agents may be unicast, multicast and broadcast messages inside the group. When an agent joins or leaves a group all the other agents inside the group are automatically notified. The main advantage of the system is that an agent can subscribe for both local and remote monitoring information and take specific actions locally and decide what has to be propagated to have a consistent view of the topology.

*2.2 Monitoring modules*

We developed a set of monitoring modules using Transaction Language 1 (TL1) (15) commands to retrieve monitoring information from the devices for two types of optical switches (Glimmerglass and Calient) and Ciena Core Director. The modules run embedded in the MonALISA service and provide near real time information for the agents. For the optical switches the system is able to monitor the optical power on ports and the state of the cross-connects inside them. For the Core Director (CD/CI) we developed modules which monitor the routing protocol (OSRP (16)) which allows us to reconstruct the topology inside the agents, the circuits (VCGs (13)), the state of cross connects, the Ethernet (ETTP/EFLOW (13)) traffic, and the allocated time slots on the SONET interfaces and the alarms raised by the CD/CI. Based on this information we have a set of filters which can take specific local actions (e.g. send emails/sms, archive the most important alarms), or notify the entire set of agents when specific conditions are met. Based on the monitoring information the agent is able to detect and to take informed decisions in case of eventual problems with the cross connections inside the switch or loss of light on the connections.

### 3. FRAMEWORK DESIGN

A schematic view of how the entire system works is presented in Fig. 2. The monitoring modules and controlling agents are distributed among the optical devices involved in the dynamic path provisioning. The status of the system is reported to the MonALISA client, where a dedicated panel was implemented especially for this purpose. In this way the user is presented at the central point with the complete status of the entire distributed system of optical devices. Even more, if the user has the correct credentials, the graphical interface provides control over the creating or removal of optical paths.

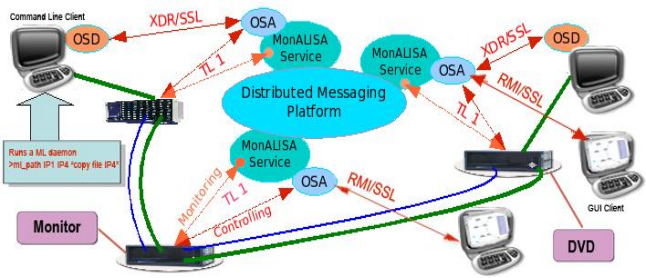

Fig. 2 System architecture

The system is integrated in a reliable and secure way with the end user applications and provides simple shell-like commands to map global connections and to create an optical path / tree on demand for any data transfer application.

*3.1 Dynamic light path circuit provisioning*

The design and implementation of the algorithm followed the main requirement for this prototype system: being able to establish an end-to-end connection in the shortest possible time. In order to achieve this, every agent in the system has the exact view of the network and adapts very quickly to changes. The network topology, implemented as a network graph [5], has agents as vertices and optical links between switches as edges, every edge having a cost associated with it. The system is modeled using a directed graph [5] because is possible to have both full-duplex and simplex links between optical switches. Every agent in the graph computes a shortest path tree using a variant of Dijkstra's algorithm. The agents system uses a two-phase commit strategy for creating the optical cross-connects and a lease mechanism to guarantee the reliability in case of partial failures.

An agent that receives a lightpath request determines, based on the local tree that is already built, if the request can be fulfilled or not. If it is possible initiates transactions with both the local and remote ports involved in the path. Once the transaction is started, the agent assigns a unique ID for the path, sends the remote cross-connect commands and after that it tries to establish the cross-connects on the local switch. An independent thread is waiting for acknowledgements from the remote agents. Any remote agent which receives such a cross-connect request starts a local transaction only with the ports involved in the cross-connect. If it succeeds in creating the cross-connect, it commits the local transaction and it sends back an "acknowledged" message, otherwise the transaction is rolled-back and a "not acknowledged" message is sent. Based on the received messages the local agent takes the decision whether or not the transaction can be committed or it has to be rolled back. The algorithm described above is reliable and guarantees that the system remains in a consistent state even if a network problem occurs. The newly created lightpath has a lease assigned which must be renewed by all the involved agents and in this way it can provide a viable mechanism for the system to recover from partial failures.

To improve the performance and the response time all the functions executed by an agent are performed in asynchronous sessions using a pool of threads. A task can be a request for a lightpath from a client, or a cross connect request coming from another agent, or a rerouting task triggered by a topology change. The only sequential part of the algorithm described above is in the "pre-commit" phase, and this involves only the ports that are supposed to be in the lightpath. Any request submitted during this phase, which do not involve these ports can be fulfilled in parallel.

Using this information, an agent is able to detect the loss-of-light on fiber, and take specific decisions if the port is part of a lightpath. The agent who detects the problem notifies the initiator, which is responsible to try to reroute the traffic through another path, if this is possible. When the initiator detects a change in topology that affects the lightpath, it forces the shortest path tree to be recalculated. Based on the new tree, the agent is able to take the decision if the light can be rerouted using other path, or it can tear down the entire path. This is very useful, because in case of successful rerouting, the problem will not disturb already established sockets, upper network layers, like TCP, not being able to detect the problem.

The routing algorithm used to establish an end to end lightpath is similar with link-state routing protocols. The work presented here uses an algorithm similar to link-state routing algorithms because they converge faster than distance-vector algorithms. In order to guarantee consistency and reliability of the entire system, a two-phase commit strategy and a lease mechanism were also developed.

*3.2 The graphical user interface*

In order to efficiently make use of the dedicated modules and agents used to monitor and control Optical Switches the MonALISA Client provides a specialized panel. This panel permits the interaction between users and the monitored Optical Switches. It offers a number of features to the end user such as complete perspective over the topology of the monitored optical networks, the possibility to monitor the state of the Optical Switches or the possibility to dynamically create new optical paths.

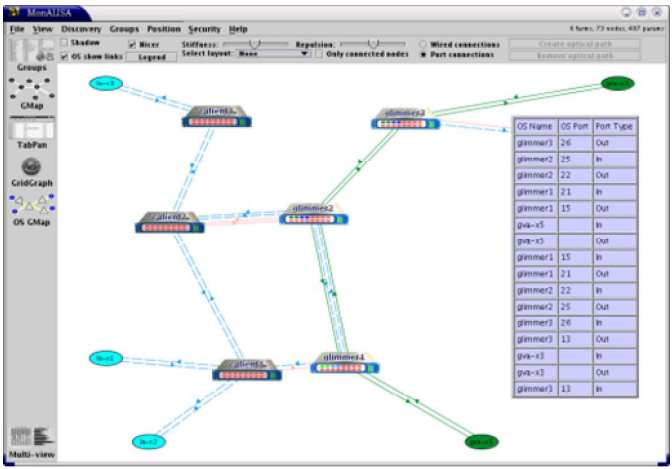

Fig. 3 Main GUI panel displays topology and optical paths

The main panel is presented in Fig. 3. The principal aspect of this panel is that it displays in an intuitive way the current topology of the monitored Optical Switches and the links between them. It also displays the active circuits, including the complete information (devices and port numbers) which are along the circuit. If the user loads a trusted certificate in the GUI and has the right permissions he can administer the remote devices (e.g. make optical path requests or manually set cross connects inside the switch) (Fig. 4)

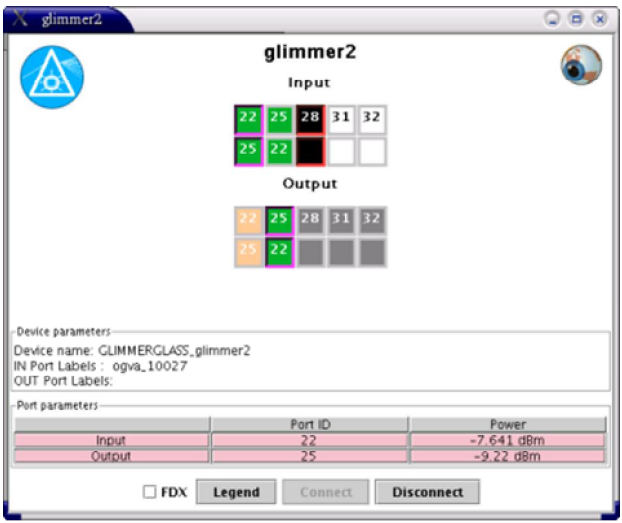

Fig. 4 Administrative panel for a particular optical switch

For the Ciena Core Director devices we developed a second panel which is able to display the OSRP topology and the virtual circuits inside the CD/CI hybrid network. The GUI can also present the status of physical links, alarms, traffic for the Ethernet flows (EFLOW) and Ethernet ports and allocated time slots for the SONET links.

*3.3 Security and user authentication*

The system is integrated in a secure and reliable way with the end used applications and provides simple shell like commands to create on demand light path between end systems. Every agent accepts light path requests through a secure connection based on X.509 certificates. The requests may be provided from a command line interface or from the MonALISA GUI Client. An optical switch daemon (OSD) must be installed on the end host machine. It locates the optical switch agents from a specific group and connects to one of them, usually to the closest one, based on topology information. The shell commands use UNIX named pipes to communicate with the daemon which routes the requests to the optical switch agent.

## 4. RESULTS

We tested our system in a real environment using two systems located at CERN (Geneva) and California Institute of Technology (CALTECH) located in Pasadena, CA, using the networking infrastructure of USLHCNet and Internet2 (17) .

For the tests we used two 10 Gbps transatlantic optical links provided by USLHCNet between CERN and Starlight (Chicago) and MANLAN (New York).On the Internet2 network, the pure optical connections are simulated using several VLANs to provide direct connections from Chicago and New York to CALTECH. The topology of the network infrastructure used is shown in Fig. 5.

The system is able to create dynamically an end to end lightpath in less than one second independent of the number of switches involved and their location. It monitors and supervises all the created connections and is able to automatically generate an alterative path in case of connectivity errors. The alternative path is set up rapidly enough to avoid a TCP timeout, and thus to allow the transfer to continue uninterrupted.

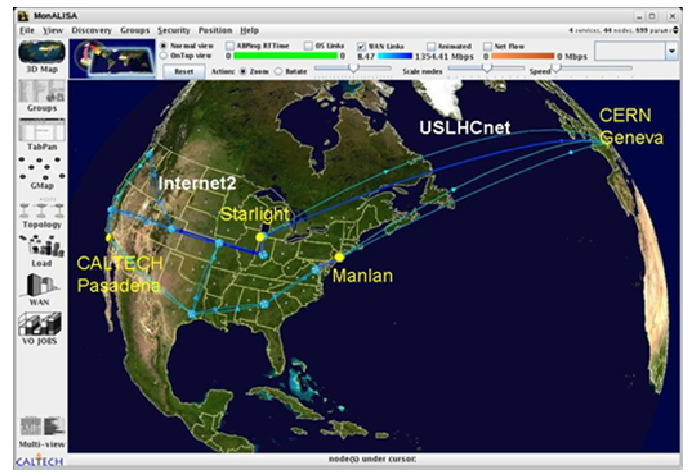

Fig. 5 USLHCNet links used for tests

In the example above a disk to disk transfer is presented, using two 4-disk servers, one at Caltech and the other one at CERN in Geneva. During the transfer four fiber cuts were simulated, corresponding to the four drops in the Fig. 6. The "fiber cut" and the reroute are done quick enough that the TCP does not sense the loss in connectivity and the transfer continues. The recovery time differs for various TCP stacks and the round trip time between end points.

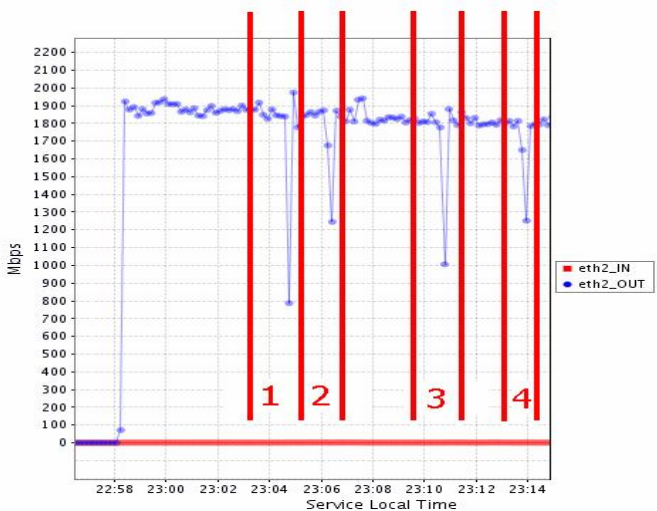

Fig. 6 FDT disk to disk transfer and rerouting

The optical fibers are simulated through two VLANs between the two optical switches. One VLAN is routed through New York and the other one through Chicago. The monitoring module is able to simulate a fiber cut. The optical agent will detect this as a real loss-of-light and will try to reroute the path.

## 5. CONCLUSIONS AND FUTURE WORK

The MonALISA agent framework was successfully used to develop an integrated network service capable to monitor, create and control dynamic circuits in hybrid networks. An end-to-end optical path was successfully tested in a real production environment. It is a distributed system, without a single point of failure. The system automatically detects network errors and is capable to create an alternative path rapidly enough to avoid a TCP timeout, so that data transport continues uninterrupted.

We are working to further develop this distributed agent system and to provide integrated network services capable to efficiently use and coordinate shared, hybrid networks and to improve the performance and throughput for data intensive grid applications. We are planning to integrate the current work in the future developments setting up dynamic VLANs using VCAT/LCAS technologies in the Ciena Core Directors.